\documentclass[prl,aps,twocolumn,floatfix,showpacs]{revtex4}

\usepackage{graphicx}
\graphicspath{{figures/}}

\usepackage{units,xspace}

\usepackage{amsmath}
\renewcommand{\vec}[1]{\boldsymbol{#1}}
\newcommand{\micron}{\ensuremath{\unit{\mu m}}\xspace}
\newcommand{\avg}[1]{\left< #1 \right>}
\newcommand{\abs}[1]{\left\vert #1 \right\vert}

\begin{document}

\title{Giant Colloidal Diffusivity on Corrugated Optical Vortices}

\author{Sang-Hyuk Lee}

\author{David G. Grier}

\affiliation{Department of Physics and Center for Soft Matter
  Research, New York University, New York, NY 10003}

\date{\today}

\begin{abstract}
A single colloidal sphere circulating around a
periodically modulated optical vortex trap can enter a
dynamical state in which it intermittently alternates between
freely running around the ring-like optical vortex and becoming 
trapped in local potential energy minima.  
Velocity fluctuations in this randomly switching state still
are characterized by a linear Einstein-like diffusion law, but
with an effective diffusion coefficient that is enhanced by
more than two orders of magnitude.
\end{abstract}

% 05.40.-a: Fluctuation phenomena, noise, and Brownian motion
% 05.60.-k: Transport processes
% 82.70.Dd: Colloids
% 87.80.Cc: Optical trapping
% 42.40.Jv: computer generated holograms

\maketitle

Brownian particles moving on tilted washboard potentials exhibit two
well-characterized limiting behaviors \cite{risken89}.  
When the
potential energy wells are deeper than the thermal energy scale,
diffusing particles become 
trapped in local minima.
Their long-time self-diffusion coefficient vanishes in this limit.
At the other extreme, tilting the washboard steeply enough to
eliminate potential energy barriers allows particles run freely
downhill.
Because
diffusion is decoupled from translation at low Reynolds numbers,
a freely running particle exhibits displacement fluctuations
characterized by its equilibrium self-diffusion coefficient,
$D_0$.
Between the trapped and running limits, particles intermittently
switch between the two states, drifting downhill at a
mean speed set by the rate at which particles are thermally
activated over barriers.
The trajectory, $x(t)$, of such an intermittently trapped
Brownian particle nevertheless is predicted
\cite{constantini99,reimann01a,reimann02a} to satisfy
the usual Einstein relation,
\begin{equation}
  \label{eq:einstein}
  \lim_{t \to \infty} \avg{x^2(t)} - \avg{x(t)}^2 = 2 D t,
\end{equation}
where $D$ is an effective diffusion coefficient.
Equally surprising
is the prediction \cite{constantini99,reimann01a,reimann02a}
$D$ can be enhanced by orders
of magnitude over $D_0$ 
at the crossover from locked to running states.

\begin{figure}[t]
  \centering
  \includegraphics[width=.9\columnwidth]{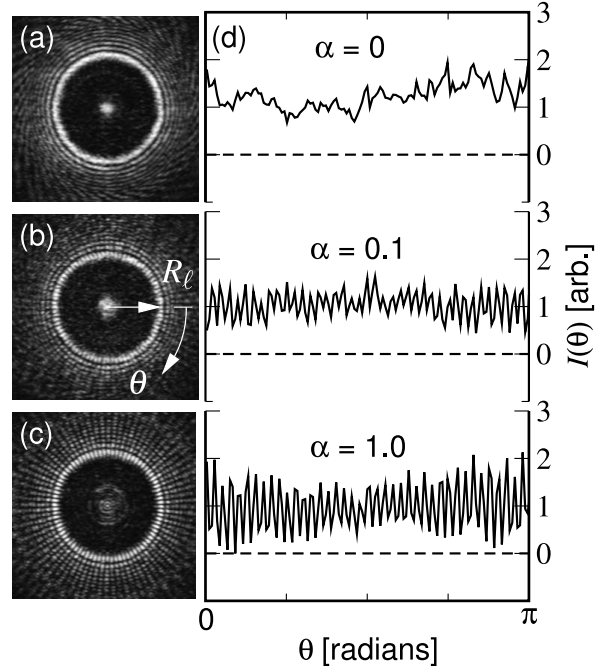}
  \caption{(a) Optical vortex with $\ell = 40$.  (b) and (c) Corrugated
    optical vortices, $\alpha = 0.1$ and 1.0, respectively.  (d)
    Circumferential intensity profiles, $I(\theta)$, measured from
    (a), (b) and (c) at the
    radius of maximum intensity, $R_\ell$.}
  \label{fig:light}
\end{figure}

In this Letter, we provide experimental confirmation
of substrate-mediated giant diffusivity
by tracking the motions of
a single colloidal sphere traveling on
tilted washboard potentials created with corrugated optical
vortex traps \cite{curtis03,guo04,lee05c}.
The optically driven particle undergoes
normal diffusion even in the intermittent regime,
with an effective diffusion coefficient that increases
more than a hundred-fold at the point of maximum
intermittency.

Our samples consist of colloidal polystyrene spheres $2a = 1.48~\micron$
in diameter 
(Bangs Laboratories, lot number 6064)
dispersed in water and confined within a glass sample
volume formed by bonding a \#1 coverslip to a
microscope slide.
This assembly is mounted on the stage of a Zeiss S100TV Axiovert
inverted optical microscope for observation.
Images are
captured by an NEC TI-324II video camera and recorded on a
Panasonic DMR-E100H digital video recorder for processing and analysis.
The polystyrene spheres sediment into a layer roughly 200~\unit{nm}
above the coverslip \cite{behrens01b}.
Individual particles are clearly resolved with a $100\times$ NA 1.4 SPlan-Apo
oil immersion objective lens that also is used to project holographic
optical traps \cite{dufresne98,curtis02,grier03,polin05,lee05c} into 
the sample.
We tracked particles' motions with 10~\unit{nm} spatial
resolution at 1/30~\unit{s} intervals using standard methods
of digital video microscopy \cite{crocker96}.
From measurements on freely diffusing spheres, we estimate 
\cite{crocker96,dufresne00} a wall-corrected self-diffusion
coefficient of $D_0 = 0.19 \pm 0.02~\unit{\micron^2/s}$.

Tilted washboard potentials were created
from superpositions of ring-like optical traps known as
optical vortices \cite{he95,simpson96,gahagan96}.
Each optical vortex in this superposition is
formed from a helical mode of light
\cite{allen92}
whose fields,
\begin{equation}
  \label{eq:helical}
  \psi_\ell(\vec{r}) = u(r) \, e^{i \ell \theta},
\end{equation}
are characterized by a radially symmetric amplitude profile, $u(r)$,
and a phase $\varphi(\vec{r}) = \ell \theta$ proportional to the
angle $\theta$ about the optical axis.
The wavefronts of such a beam take the form of an $\ell$-fold helix
whose pitch determines the radius, $R_\ell$, 
of the projected ring of light \cite{curtis03,sundbeck05}.
A typical optical vortex with $\ell = 40$ and $R_\ell = 4.2 \pm 0.1~\micron$
appears in Fig.~\ref{fig:light}(a).
The focused ring of light acts like an optical gradient force trap, drawing
nearby dielectric particles to its circumference.
The helical pitch also endows each photon 
in the beam with $\ell \hbar$ orbital angular momentum \cite{allen92}
that can be transferred to an illuminated object.
The resulting torque causes a trapped object
to circulate around the ring \cite{he95a}.

Superposing optical vortices with opposite helicities, $\ell$ and $-\ell$,
creates corrugated optical vortices such as the examples in
Figs.~\ref{fig:light}(b)
and (c) whose circumferential profiles are sinusoidally
modulated \cite{guo04}
with $2 \ell$ intensity maxima \cite{lee05c}.
An even superposition creates a so-called optical cogwheel
\cite{jesacher04} consisting of bright spots arranged in a
circle of radius $R_\ell$.  This superposition carries no net orbital
angular momentum and thus exerts no torque.
A more general superposition,
\begin{equation}
  \label{eq:superposition}
  \psi(\vec{r}) = 
  \frac{\psi_\ell(\vec{r}) + 
    \sqrt{\alpha} \, \psi_{-\ell}(\vec{r})}{
    \sqrt{1 + \alpha}},
\end{equation}
still exerts a torque, but also has a sinusoidal
corrugation whose depth is set by $0 \le \alpha \le 1$.
The two limits, $\alpha = 0$ and $\alpha = 1$, 
correspond to an ideal optical vortex and an optical cogwheel,
respectively.
We vary $\alpha$ by calculating phase-only holograms encoding the
desired superposition using the direct search algorithm
\cite{polin05} and projecting the results with a Hamamatsu
X7550 PAL-SLM spatial light modulator.
Our experiments were performed with 1.5~\unit{W}
of laser light at a wavelength of $\lambda = 532~\unit{nm}$ from
a Coherent Verdi laser.

The images of corrugated optical vortices in Fig.~\ref{fig:light}
were captured by placing a
mirror in the microscope's focal plane and capturing the
reflected light with the objective lens.
Circumferential
intensity profiles, $I(\theta) = \abs{\psi(R_\ell,\theta)}^2$
measured from these images are plotted
in Fig.~\ref{fig:light}(d) and
reveal errors in $\alpha$ smaller than 5 percent.
Additional intensity
variations of roughly 10 percent arise from imperfections in the optical
train and so are independent of $\alpha$.
These variations have an even smaller effect on the potential
energy landscape experienced by the particle because the particle's
finite extent tends to smooth them over \cite{ladavac04,pelton04a}.

A photon in the superposed beam has probability $1/(1+\alpha)$ to
have orbital angular momentum $+\ell\hbar$ and probability
$\alpha/(1+\alpha)$
to have orbital angular momentum $-\ell\hbar$.
The corrugated optical vortex therefore carries a local
orbital angular momentum flux
$(\ell \lambda/c) \, [(1 - \alpha)/(1 + \alpha)] \,
  I(\theta)$,
where $c$ is the speed of light.
A fraction of this orbital angular momentum
is transferred to a trapped object, and drives it around the ring.
Circumferential intensity gradients,
$\partial_\theta I(\theta)$,
modulate this torque, and
also induce optical gradient forces.
The overall circumferential force
therefore has the form
\begin{equation}
  \label{eq:force2}
  F(\theta) = F_0 \, \left[A(\alpha) + 
    B(\alpha) \, \cos(2 \ell \theta) +
    \eta(\theta) \right],
\end{equation}
with $A(\alpha) = (1-\alpha)/(1 + \alpha)$ and 
$B(\alpha) = 2 \, \sqrt{\alpha (A^2 + \xi^2)} / (1+\alpha)$.
The force scale, $F_0$ is proportional to the
laser beam's power and
depends on the particle's shape, size and
composition
\cite{ladavac04,pelton04a}.
Given this, the washboard's tilt depends only 
on the excess, $A(\alpha)$ of right-helical
photons in the corrugated optical vortex.
The sinusoidal term, by contrast,
includes a material- and geometry-dependent constant, $\xi$, accounting
for the relative strength of the optical gradient force.
We have omitted an irrelevant phase from the sinusoidal term's
argument, and will ignore
landscape's roughness, $\eta(\theta)$,
in what follows.

\begin{figure}[t]
  \centering
  \includegraphics[width=\columnwidth]{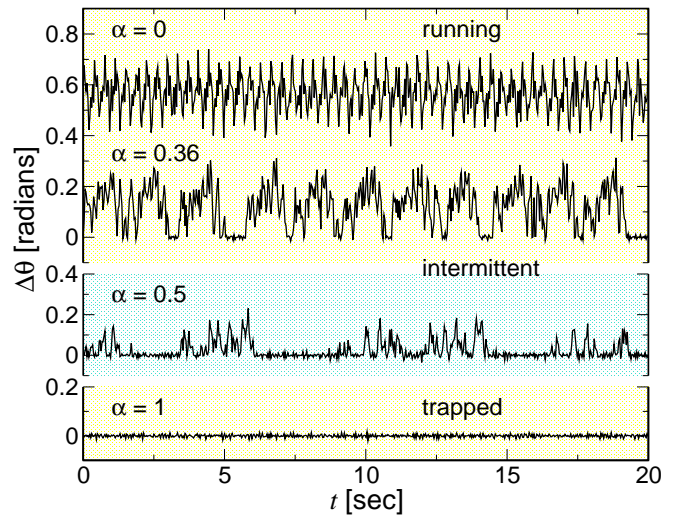}
  \caption{Instantaneous angular speed for a single 1.48~\micron
    diameter polystyrene particle
    circulating around corrugated optical vortices at $\ell = 40$,
    $P = 1.5~\unit{W}$ and $\alpha = 0$, 0.36, 0.5 and 1.0.}
  \label{fig:dtheta}
\end{figure}

Figure~\ref{fig:dtheta} shows brief samples from typical 
single-particle trajectories
in the trapped, running and intermittent regimes.
Here, we have plotted the angular displacement, $\Delta \theta(t) =
\theta(t + \delta t) - \theta(t)$ over the period
$\delta t = 1/30~\unit{s}$ of one video frame.
The driving term vanishes
in the cogwheel limit, $\alpha = 1$, 
and the particle remains trapped in a
single local minimum of the potential, where it undergoes thermally driven
fluctuations about its equilibrium position.
In the freely running limit, $\alpha = 0$, it circulates around
the ring nearly three times a second.
Periodic features in the running state's displacements
result from the particle passing repeatedly over the
disordered landscape,
$\eta(\theta)$.
From these, we
estimate $\avg{\abs{\eta(\theta)}^2} \approx 0.01$.
At intermediate values of $\alpha$, the particle makes thermally
activated transitions between trapped and running states so that
its trajectory is characterized by intermittent bursts of motion
resembling random telegraph switching noise.

\begin{figure}[t]
  \centering
  \includegraphics[width=\columnwidth]{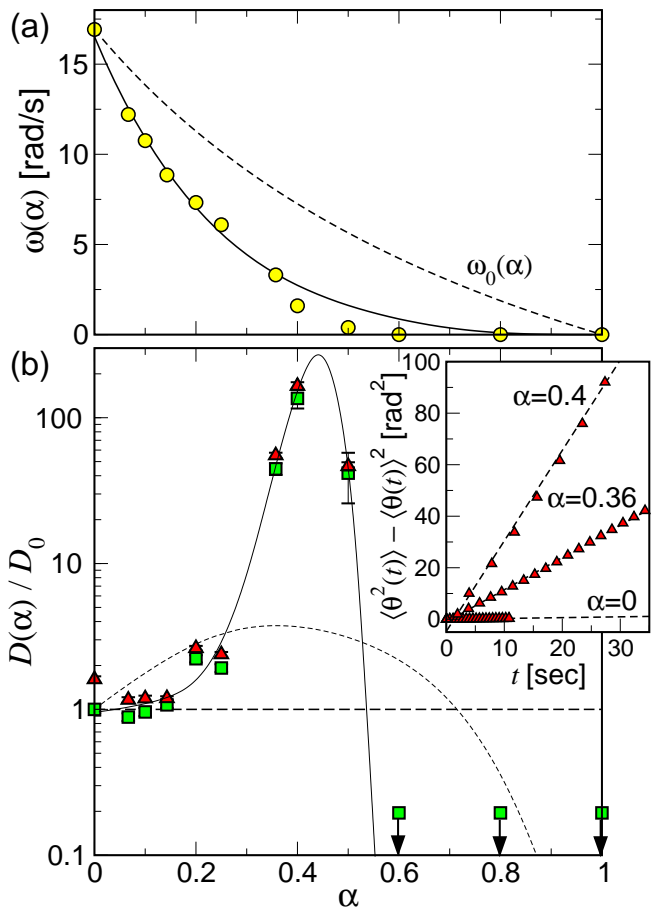}
  \caption{(a) Dependence of circulation frequency on depth of
    corrugations.  The solid curve is a fit to
    Eq.~(\ref{eq:stratonovich}) yielding $\xi = 0.19$ and
    $F_0 = 1.5~\unit{pN}$.  The dashed curve is an estimate for
    $\omega_0(\alpha)$ from Eq.~(\ref{eq:force2}).
    (b) Diffusion coefficients estimated from the inset plots of
    mean-squared displacements (triangles)
    show a
    dramatic enhancement around $\alpha = 0.4$.  Squares denote
    equivalent results obtained from Eq.~(\ref{eq:perioddiffusion}).  
    Long-term
    self-diffusion coefficients are unmeasurably small in the trapped state for
    $\alpha \ge 0.6$.  The solid curve is a
    guide to the eye.  The dashed curve is a comparison to 
    Eq.~(\ref{eq:reimann}) using parameters from (a).
    }
  \label{fig:diffusion}
\end{figure}

Averaging $\Delta \theta(t)$ over a period long compared with the trajectory's correlation
times yields an estimate for the mean circulation rate
$\omega(\alpha)$, plotted in Fig.~\ref{fig:diffusion}(a).
The net orbital angular momentum flux driving this circulation
decreases as $\alpha$ increases, and we estimate the free circulation
rate to be $\omega_0(\alpha) = A(\alpha) \, \omega(0)$
from Eq.~(\ref{eq:force2}).  This is drawn as a dashed curve in
Fig.~\ref{fig:diffusion}(a).
The solid curve in Fig.~\ref{fig:diffusion}(a) is a
comparison to
Stratonovich's exact expression for the mean drift velocity
\cite{reimann01a,reimann02a}, 
\begin{align}
  \omega(\alpha) & = 
  \frac{1 - e^{- 2 \pi \beta R_\ell F_0 A(\alpha)}}{
    R_\ell \, \int_0^{2\pi} I_+(\theta) \, \frac{d\theta}{2\pi}}, 
  \quad \text{where} 
  \label{eq:stratonovich}\\
  I_\pm(\theta) & = \frac{R_\ell}{D_0} \, e^{\pm \beta V(\theta)} 
  \int_0^{2\pi} e^{\mp \beta V(\theta \mp \phi)}
    e^{- \beta R_\ell F_0 A(\alpha) \phi} \, d\phi \nonumber
\end{align}
and where 
$V^\prime(\theta) = F_0[B(\alpha) \cos(2\ell\theta) + \eta(\theta)]$.

Slow instrumental drifts in $\omega(\alpha)$ amounting to
a few percent over several minutes 
can be estimated \cite{chatfield03} and subtracted off to reveal
the linear growth of mean-squared positional fluctuations shown in 
the inset to Fig.~\ref{fig:diffusion}(b).
The associated effective diffusion coefficients, $D(\alpha)$ are
plotted as triangles in Fig.~\ref{fig:diffusion}(b).
These values agree well
with those obtained \cite{reimann01a,reimann02a}
from fluctuations in the time, $T_j(\alpha)$, required to complete 
the $j$-th circuit,
\begin{equation}
  \label{eq:perioddiffusion}
  D(\alpha) = 2 \pi^2 R_\ell^2 \, 
  \frac{\avg{T_j(\alpha)^2}_j - \avg{T_j(\alpha)}_j^2}{\avg{T_j(\alpha)}_j^3},
\end{equation}
which are plotted as squares.

The measured effective diffusion coefficient agrees with
the equilibrium value, $D(\alpha) = D_0$ when the particle
is in the free-running state
$\alpha < 0.2$.
Larger values of $\alpha$ correspond to
deeper corrugations that
tend to trap the particle for longer periods.
Longer periods of localization
might be expected to reduce the particle's 
effective diffusion
coefficient.
Indeed $D(\alpha) = 0$ when the particle
is trapped altogether for $\alpha \ge 0.6$.
Instead, intermittent trapping dramatically increases the
effective diffusion coefficient, with $D$ exceeding
$100~D_0$ at $\alpha = 0.4$.

This extraordinary substrate-mediated enhancement
of the effective diffusivity is accounted for by the
exact formulation, analogous to Eq.~(\ref{eq:stratonovich}),
due to Reimann \emph{et al.} \cite{reimann01a,reimann02a},
\begin{equation}
  \label{eq:reimann}
  D = D_0 \, 
  \frac{\int_0^{2\pi} I_+^2(\theta)I_-(\theta) \, \frac{d\theta}{2\pi}}{
    \left[\int_0^{2\pi} I_+(\theta) \, \frac{d\theta}{2\pi}\right]^3}.
\end{equation}
Analysis of Eq.~(\ref{eq:reimann}) reveals that
peak diffusivity should occur when the driving force is
just barely balanced by the periodic modulation.
Neglecting $\eta(\theta)$, this occurs when $\alpha$
satisfies $A(\alpha) = B(\alpha)$.
Estimating the peak position to be
$\alpha = 0.41 \pm 0.01$, we obtain
$\xi = 0.19 \pm 0.01$.
The sharp peak in $D(\alpha)$ therefore
can be used to probe how
absorption and scattering transfer
orbital angular momentum to objects
trapped in optical vortices.
The form for $D(\alpha)$ predicted by Eq.~(\ref{eq:reimann})
is plotted as the dashed curve in Fig.~\ref{fig:diffusion}(b).
Whereas Eq.~(\ref{eq:stratonovich}) agrees well with $\omega(\alpha)$,
Eq.~(\ref{eq:reimann}) describes a much broader and weaker 
peak in $D(\alpha)$
than is observed experimentally.
It appears, therefore, that $D(\alpha)$ is exceptionally sensitive
to details of the tilted washboard potential.

\begin{figure}[t]
  \centering
  \includegraphics[width=\columnwidth]{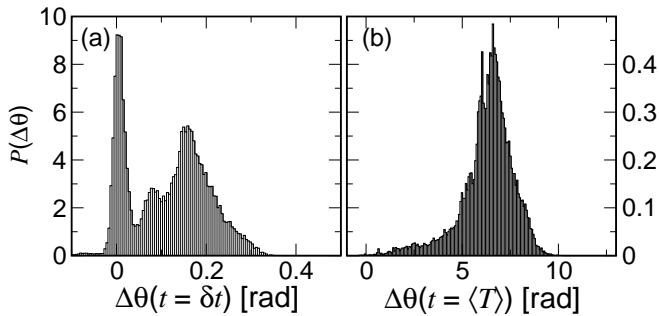}
  \caption{Displacement histograms in the intermittent regime
    at $\alpha = 0.36$.
    (a) Short time, $t = 33~\unit{ms}$.  (b) Long-time, 
    $t = \avg{T} = 1.3~\unit{s}$.}
  \label{fig:histogram}
\end{figure}

Histograms of angular displacements in Fig.~\ref{fig:histogram}
provide insights into this sensitivity.
A freely running particle's
displacements fall into a nearly Gaussian distribution,
whose width and peak position both increase linearly in time.
In the intermittent state, by contrast, 
particles spend much of their time localized in traps,
so that the short-time displacement probability is highly
non-Gaussian, as shown in Fig.~\ref{fig:histogram}(a).
Because the potential energy landscape is periodic
and the effective particle density is
fixed, the displacement probability must evolve into
a Gaussian distribution through the central limit theorem.
This self-averaging can be effective over as little as
a single mean first-passage time $\avg{T}$, as shown in 
Fig.~\ref{fig:histogram}(b).
This accounts for the essentially normal diffusion evident
in the inset to Fig.~\ref{fig:diffusion}(b).

The peaked structure of the probability distribution in 
Fig.~\ref{fig:histogram}(a) also suggests a qualitative
explanation for the overall enhancement of diffusivity.
A particle undergoing intermittent transport has a probability
$p(t)$ to be trapped for time $t$,
and a probability $1 - p(t)$ to travel with
an angular speed $\omega_0$ set by the washboard's overall tilt.
The mean drift speed in this highly simplified two-state
model is $\omega = (1-p) \, \omega_0$.
The running state is characterized by thermal
fluctuations in the mean-squared angular speed of
magnitude $D_0 / (R_\ell^2 t)$.
This, however, can be dominated by
fluctuations due to thermally activated 
transitions between the
stationary and running
states.
Taking $t = \pi/(\ell \omega_0)$ to be the time required to
travel between potential wells in the running state,
\begin{align}
  D & \equiv \frac{R_\ell^2 t}{2} \, \left( \avg{\omega^2} -
    \avg{\omega}^2\right) \\
  & = \frac{p(1-p)}{2} \, 
  \frac{\pi R_\ell^2}{\ell} \, \omega_0 + (1 - p) \, D_0.
\end{align}
The effective diffusion coefficient therefore can be made
arbitrarily large by increasing $\omega_0$,
with an upper limit set by the onset of inertial effects.

Giant diffusivity can degrade the performance of
sorting methods such as gel electrophoresis and optical
fractionation that exploit differential transport through a
structured medium.
In particular, the relative distribution $\Delta x/L$
of a sample that has traveled a mean distance $L$ through the
landscape at speed $v$ is $\sqrt{2 D / (L v)}$, 
which can diverge with the effective diffusion coefficient, $D$.
This effect may be responsible for anomalous band
broadening in electrochromatography \cite{rebscher94}.
Figure~\ref{fig:diffusion}(b) demonstrates, however, that undesirable
dispersal due to giant diffusivity can be overcome by more
rapid driving, and that a small increase in driving force can have a
disproportionately large effect on sorting resolution.
On the other hand, substrate-mediated giant diffusivity 
should be useful for thoroughly
mixing and dispersing materials in microfluidic environments, and
might also provide a strategy for enhanced mixing in granular materials.

We have benefited from conversations with Yael Roichman.
This work was supported by the National Science Foundation through
grant number DMR-0451589.  SL acknowledges support of a Kessler Family
Foundation fellowship.

%\bibliographystyle{apsrev}
%\bibliography{abbreviations,grier,tweezer,dgg,giant}

\end{document}